\documentclass[osajnl,letter,twocolumn,showpacs,superscriptaddress,10pt]{revtex4-1} 
\usepackage{amsmath,amssymb}

   \usepackage{graphicx}

\begin{document}

\title{ Noisy metamolecule: strong narrowing of fluorescence line }

\author{E.~S.~Andrianov}
\affiliation{All-Russia Research Institute of Automatics, 22 Sushchevskaya, Moscow 127055, Russia}
\affiliation{Department of Theoretical Physics, Moscow Institute of Physics and Technology, 141700 Moscow, Russia}

\author{N.~M.~Chtchelkatchev}
\affiliation{Department of Theoretical Physics, Moscow Institute of Physics and Technology, 141700 Moscow, Russia}
\affiliation{Department of Physics and Astronomy, California State University Northridge, Northridge, CA 91330}
\affiliation{L.D. Landau Institute for Theoretical Physics, Russian Academy of Sciences, 142432 Chernogolovka, Russia}

\author{A.~A.~Pukhov}
\affiliation{All-Russia Research Institute of Automatics, 22 Sushchevskaya, Moscow 127055, Russia}
\affiliation{Department of Theoretical Physics, Moscow Institute of Physics and Technology, 141700 Moscow, Russia}
\affiliation{Institute for Theoretical and Applied Electromagnetics, 13 Izhorskaya, Moscow 125412, Russia}

\begin{abstract}
We consider metamolecule consisting of bosonic mode correlated with the two-level system: it can be, for example, plasmonic mode interacting with the quantum dot.  We focus on the parameter range where all the correlations are strong and of the same order: interaction between bosonic mode correlated with the two-level system, external coherent drive  and dissipation. Quantum Monte-Carlo simulations show a  fluorescence of this system at dissipation larger than the driving amplitude and strong (by the order of magnitude) narrowing of its spectral line. This effect may be related to kind of a quantum stochastic resonance.  We show that the fluorescence corresponds to finite domain over the coherent drive with sharp lower threshold and there is splitting of the Wigner function.
\end{abstract}

\maketitle

Correlated bosonic modes and discrete spectrum systems   attract much attention last time: they are key elements of applications in quantum optics~\cite{carmichael2003Nature,robins2008Nature,Liu2007PhysRevLett,Liu2009PhysRevLettPhotons,dubin2010Nature},  chemistry~\cite{jain2008noble}, and biology~\cite{huang2010gold} that includes nanolasers~\cite{noginov2009demonstration,oulton2009Nature,stockman2010spaser,lu2012Science,bergman2003surface,Richter2015PhysRevB}, qubits, quantum shuttles~\cite{Novotny2004PhysRevLett}, cold atoms~\cite{Liu2009PhysRevLett}, plasmonic nanoprobes and biosensors.~\cite{Maier2007} Today technology enables to reduce the size of the bosonic mode host up to nanometer scale. In particular, such plasmonic nanosystems are often referred to as ``metamolecules''~\cite{Fedotov2010PhysRevLett,shafiei2013Nature,Maple2014Opt.Express,Gregory2014NewJPhys,kuzyk2014NatureMaterials,Khai2015JAP}. Here we consider as metamolecule also plasmonic nanosystems interacting with quantum dot~\cite{Ridolfo2010PhysRevLett}.

In most applications, correlated bosonic modes and discrete spectrum systems in the first approximation may be reduced to the Jaynes-Cummings class of models. They are quite well investigated  except the parameter range where there is  no well defined small or large interaction parameter or there is high level of dissipation and there are large fluctuations. After Ref.~\cite{carmichael1999book1} we refer to these range as to the ``no-man's-land''~\cite{li2013ice}.  In these cases advanced computational efforts are required~\cite{hughes2011PRL,Dombi2014JPB}. For applications mentioned above this parameter range is quite typical. In the real systems the noise is always present. Usually this is the problem especially  for applications related to quantum effects. Recent experiments on single DNA hairpins show amplification of the response by stochastic noise~\cite{SILCHENKO2003,Ritort2012PRX}.  Here we show partially motivated by experiments that noise may effectively increase coherence in metamolecule while the system occupies the ``no-man's-land''.

\begin{figure}[b]
  \centering
  \includegraphics[width=0.30\columnwidth]{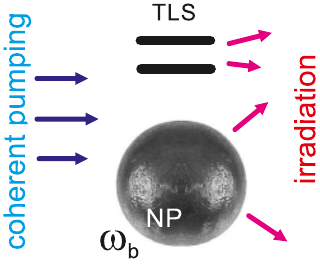}\\
   \caption{(Color online) Possible realisation of noisy metamolecule consisting of plasmon in metallic nanoparticle~\cite{bergman2003surface,noginov2009demonstration,stockman2010spaser} (or nanorod~\cite{oulton2009Nature,lu2012Science}) correlated with the two-level system  (atom, molecule or a quantum dot).}\label{figdevice}
\end{figure}
In the simplest case the Jaynes-Cummings model (JCM) corresponds to the noisy bosonic mode correlated with the two-level system~\cite{scully1997quantum,walls2008quantum}. Electromagnetic mode in the resonator or the plasmon in the nanostructure may serve as the bosonic mode while ground and excited states of  a molecule or a quantum dot may serve as the two-level system~\cite{carmichael2003Nature,robins2008Nature,dubin2010Nature,yasir2015tunable}. In the ``no-man's-land'' the Rabi frequency is of the order of dissipation in bosonic mode. From the fluctuation-dissipation theorem dissipation means noise.  We do nonequilibrium quantum Monte-Carlo  simulations~\cite{Dalibard1992PRL,Breuer2002book,hughes2011PRL} of the system and show that  under coherent drive there is a kind of a dynamical ``phase transition'' driven by noise accompanied by by the narrowing of the fluorescence spectral line by the order of magnitude, enhancement of the signal to noise ratio  and  splitting of the Wigner function --- it plays the role of the effective ``order parameter''.  From experimental point of view this fluorescence regime may find wide applications where the strong noise is the unavoidable problem.

We consider within JCM metamolecule, see Fig.~\ref{figdevice}. The dynamics of the system in Markov approximation  is governed by the Lindblad equation~\cite{Breuer2002book}:
\begin{multline}\label{2}
d\hat \rho /dt =  - i\left[ {\hat H,\hat \rho } \right] + \frac{{{\gamma_b}}}{2}\left( {2\hat a\hat \rho {{\hat a}^+ } - \hat \rho {{\hat a}^+ }\hat a - {{\hat a}^+ }\hat a\hat \rho } \right) +
\\
 +\frac{{{\gamma_a }}}{2}\left( {2\hat \sigma \hat \rho {{\hat \sigma }^+ } - \hat \rho {{\hat \sigma }^+ }\hat \sigma  - {{\hat \sigma }^+ }\hat \sigma \hat \rho } \right),
\end{multline}
where ${\gamma_a}$ and ${\gamma _b}$ are the dissipation rates of the two-level system and bosonic mode correspondingly and
$
\hat H = {\hat H_{b}} + {\hat H_{a}} + {\hat V_{a - w}} + {\hat V_{b - w}} + {\hat V_{a - b}}.
$
Here ${\hat H_{b}} = {\omega _{b}}{\hat a^ + }\hat a$ is the Hamiltonian of the bosonic mode ($\hat a$ is annihilation operator), ${\hat H_{a}} = {\omega_{a}}{\hat \sigma^+ }\hat \sigma $ is a two-level atom Hamiltonian ($\hat \sigma  = \left| g \right\rangle \left\langle e \right|$ is operator of transition from the excited state $\left| e \right\rangle $ to the ground one $\left| g \right\rangle $), ${\hat V_{b-a}} = {\Omega_R}({\hat a^+ }\hat \sigma  + {\hat \sigma^+ }\hat a)$ is the interaction between the bosonic mode and the two-level system (TLS) where ${\Omega_R}$ is the Rabi frequency.  The coherent drive of TLS and bosonic mode is described by ${\hat V_{a-w}} =\Omega_a \left( {\hat \sigma  + {{\hat \sigma }^+ }} \right)$ and ${\hat V_{b-w}} =  \Omega_b ({\hat a^+ } + \hat a)$. We choose the system of units here and below where $\hbar=1$.

Eq.~\eqref{2} has been intensively investigated last decades both analytically and numerically. However this is not so for the ``no-mans-land'', where any analytical progress is very hard. (In~\onlinecite{hughes2011PRL} ``no-man's-land" has been investigated but with slightly different model and parameter range.) In addition, in this parameter region the system is in between quantum and classical regimes since the average boson number,  $N=\langle \hat a^\dag \hat a\rangle$, is larger than unity but not big enough for applicability of the semiclassical methods or $1/N$ expansion~\cite{carmichael2008book2}. Note that experimentally metamolecule is realized in nanolaser type systems.
\begin{figure}[b]
  \centering
  \includegraphics[width=0.9\columnwidth]{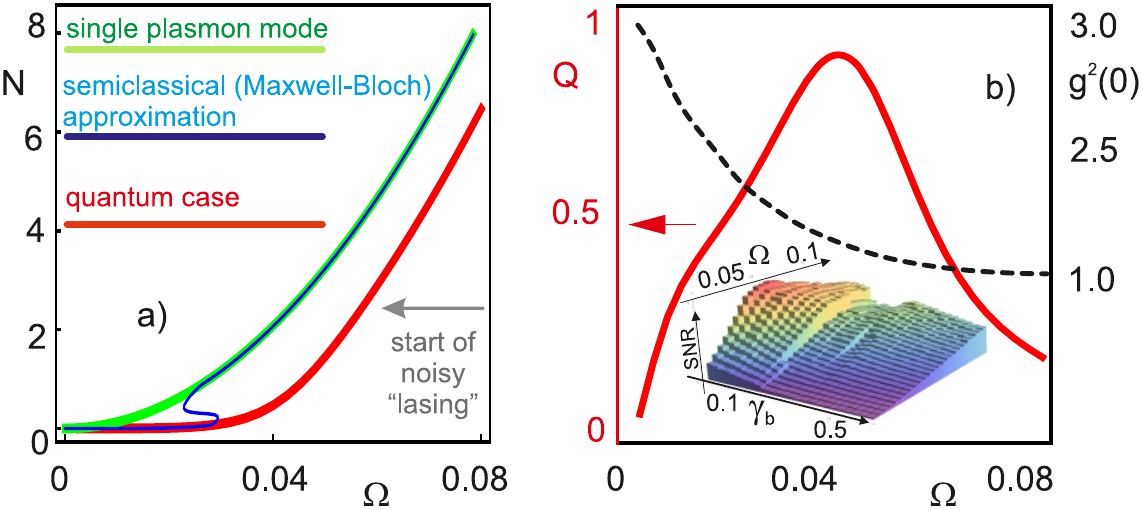}\\
  \caption{(Color online) ``Lasing" threshold. a)  boson number $N$, b) Mandel parameter (red solid) and coherence function (black dash). Maximum corresponds to the threshold where narrow line fluorescence appears. Insert shows in arbitrary units the signal to noise ratio (SNR). It is inversely proportional to the boson mode linewidth. }\label{figsemiclassics}
\end{figure}

\begin{figure}[t]
  \centering
  \includegraphics[width=0.95\columnwidth]{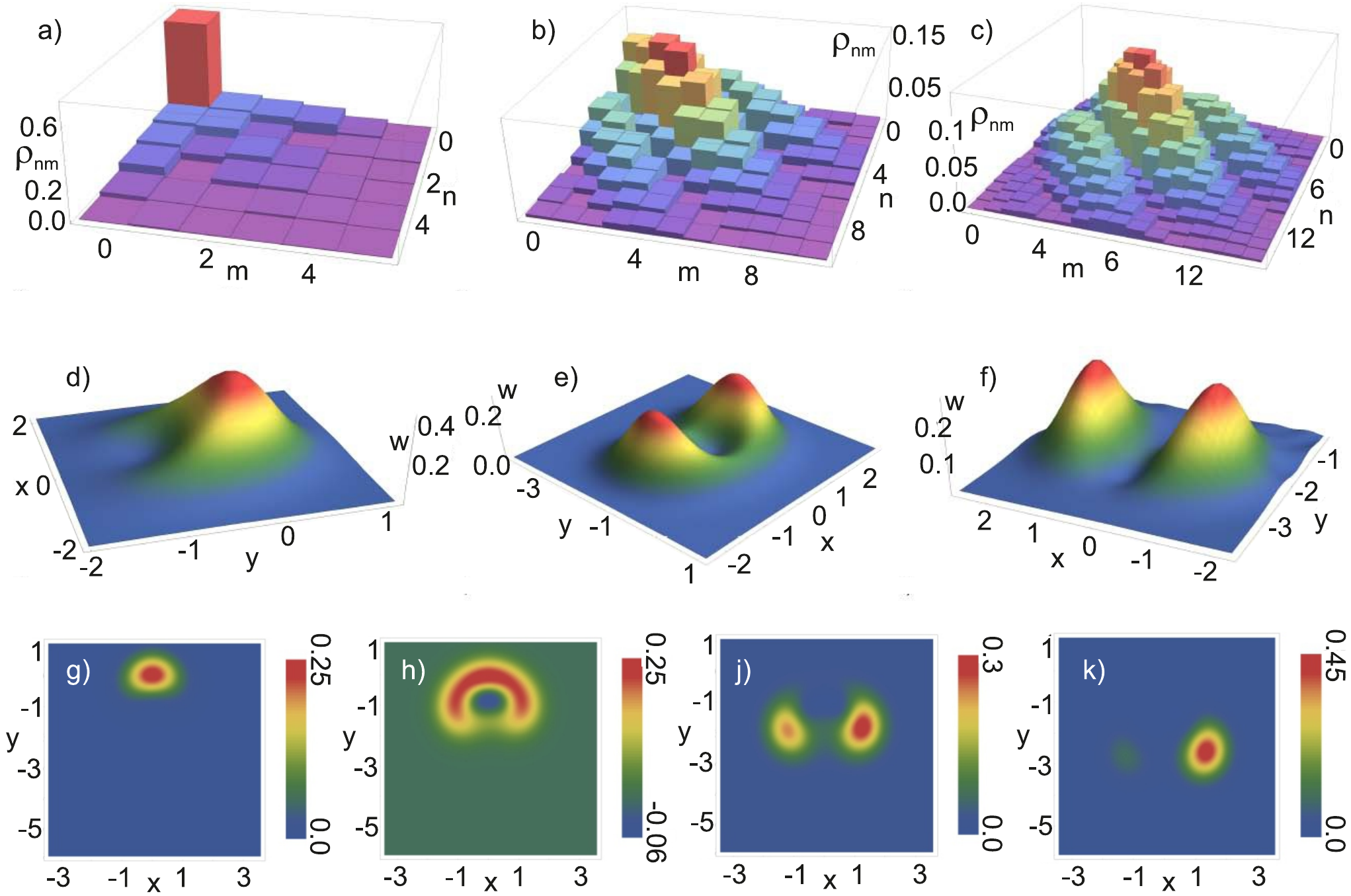}\\
  \caption{(Color online)  ``Lasing" threshold and above. (a) -( c) The steady-state reduced bosonic density matrix in the basic of the occupation numbers. Its multimodality  corresponds to the narrow line fluorescence regime (``lasing''). (d) -( f) show the Wigner function where $x$ and $y$ are real and imaginary part of the coherent state. Break up of the Wigner function into two peaks corresponds to narrow line fluorescence regime. Graphs (a),(d); (b),(e); and (c),(f) were prepared at $\Omega=0.04, 0.06,  0.08$. Density plots (bootom raw) show the Wigner function at (g) $\Omega=0.03$, (h) $\Omega=0.05$ (In the region at the center of the ``attol'' the Wigner function is negative, i.e. this is non-classical state),  (j) $\Omega=0.07$ and (k) $\Omega=0.09$.}\label{Distr}
\end{figure}

In Ref.~\cite{noginov2009demonstration} the plasmonic nanolaser has been experimentally realised.  This experiment is realisation of JCM where the role of the bosonic mode plays  the plasmon in the metallic nanoparticle, see Fig.~\ref{figdevice}, while TLS correspond to quantum dots. For this system ${\gamma _b} \sim {10^{12}} - {10^{14}}{s^{ - 1}}$. Due to strong field localization in the plasmonic mode the Rabi interaction constant may become comparable with the damping in the plasmonic mode~\cite{klimov_book_nanoplasmonics}.  ${\Omega _R} \sim {10^{12}} - {10^{13}}{s^{ - 1}}$. We should note that in this experiment many TLS have been correlated with plasmonic bosonic mode and in our model --- only one. The other difference is the external drive: in the experiment the system was driven by 5ns pulses while we consider coherent drive.  We do not see however any problem with the realisation of  coherent drive by an external laser and the use of a single quantum dot. The realistic amplitude of the laser field is $\sim10^3-10^4 V/m$. For realistic semiconductor quantum dots typical relaxation rate is ${\gamma _a} \sim 10^{9}-10^{11} s^{-1}$~\cite{noginov2009demonstration}. Another possibility is quantum dot or molecule in the cavity~\cite{carmichael2003Nature,robins2008Nature,dubin2010Nature,yasir2015tunable} like we  sketch in Fig.~\ref{figdevice}a. More recent realization corresponds to biology where coherent effects in the presence of large noise have been observed~\cite{romero2014quantum}.

\begin{figure*}[t]
  \centering
  \includegraphics[width=\textwidth]{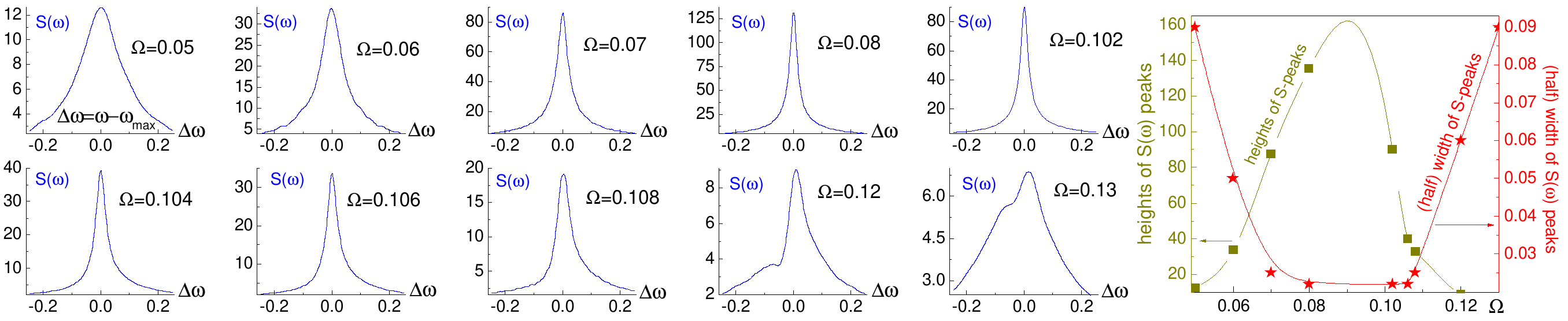}\\
  \caption{(Color online) Spectra evolution of the bosonic mode: $\Omega=0.05, 0.06,  0.07, 0.08, 0.102, 0.104, 0.106, 0.108, 0.12, 0.13$.   }\label{spectrum}
\end{figure*}
So, we imply the following values of constants: ${\gamma _a}=10^{11} s^{-1}$ and  ${\Omega _R}=10^{13} s^{-1}$. The dumping rate of bosonic mode ${\gamma _b}\in(0.5\cdot10^{13} s^{-1},\,4\cdot10^{13} s^{-1})$ which corresponds to different nanocavity materials. Also we choose $\Omega_b=10\Omega_a\equiv\Omega$ which corresponds to more intensive interaction between external field with nanocavity.  It takes place when plasmonic mode is excited, e. g., see Ref.~\onlinecite{noginov2009demonstration}.  The absolute value of interaction constant $\Omega$ with external field we will vary from $5\cdot10^{12} s^{-1}$ to $1\cdot10^{13} s^{-1}$ which corresponds to experimentally achievable laser field amplitude,  $\simeq 10-100 V/m$. In simulation we normalize all values on the $10^{-14}s$. So we consider the ``no-mans-land'' case when ${\gamma _a} \simeq {\Omega _R} \simeq {\Omega} \gg {\gamma _\sigma }$. Also we suppose the exact resonance between bosonic mode and two-level system.

What happens beyond these parameters? In the insert of Fig.~\ref{figsemiclassics}b we show the signal to noise ratio; in fact, any point within its broad extremum ridge corresponds to narrow line fluorescence.

To solve Eq.~\eqref{2} we use  the quantum Monte Carlo algorithm where the Lindblad equation reduces to the piecewise deterministic process on  quantum trajectories~\cite{Dalibard1992PRL,Breuer2002book,hughes2011PRL}, with the Gilbert space dimension  $N\approx 100$; this is more than enough, see Fig.~\ref{Distr}(a)-(c), where the peaks show the most probable $N$.

First we investigate the evolution of the average boson number $N$ with the drive amplitude $\Omega$  to detect the parameter range where $N$ grows that might correspond to narrow line fluorescence. We plot $N$ in Fig.~\ref{figsemiclassics}a not only in  the quantum case, corresponding to the exact solution of the master equation~\eqref{2}, but for comparison we show in Fig.~\ref{figsemiclassics}a results of well-known approximate solutions for $N$: 1) the semi-classical approximation and 2) ``noninterracting'' case corresponding to the dynamics of the bosonic mode in the external field but without interaction with the TLS.

The  semiclassical solution of Eq.~\eqref{2} implies total neglecting of quantum fluctuations. As follows from  Fig.~\ref{figsemiclassics}a there is hysteresis (bistability) in this case. Exact solution (red curve) of~\eqref{2} shows no hysteresis: it disappears due to strong quantum fluctuations~\cite{carmichael1999book1}. Outside the bistability the value of population inversion, $\langle[\hat\sigma^+,\hat\sigma]_-\rangle>0$. So TLS can be pumped using just a monochromatic external wave and dissipative bosonic mode in agreement with the  ``bad-cavity limit''~\cite{Cirac1992PRA}.

In our problem quantum fluctuations play important role as shows the Mandel parameter $Q=(\langle \hat{N}^2 \rangle - \langle \hat{N} \rangle^2)/\langle \hat{N} \rangle -1 = \langle \hat{N} \rangle (g^{(2)}(0)-1 )$, where $\hat N=\hat a^+\hat a$ and $g^{(2)}(\tau)=\langle { \hat a^+(t+\tau)} { \hat a^+(t+\tau)} { \hat a (t)} { \hat a (t)} \rangle$.  $Q$ has maximum at the driving field amplitude corresponding to the beginning of narrow line fluorescence regime, see Fig.~\ref{figsemiclassics}b.

Fig.~\ref{figsemiclassics} shows linear growth of $N$ above certain threshold  of drive. Below we investigate the parameter range of the transition. In Figs.~\ref{Distr} (a)-(c) we present the reduced density matrix  in the  occupation number representation for different values of the coherent field amplitude $\Omega$. At small $\Omega$, much below the transition to narrow line fluorescence, there is one pronounced peak in the density matrix   diagonal corresponding  to $N=0$. At the transition and above the density matrix becomes essentially nondiagonal. We see  the appearance of the phase multimodality of the matrix elements in the narrow line fluorescence regime: i.e., splitting of the density matrix distribution. In Figs.~\ref{Distr} (a)-(c) the density matrix splits at least on three ridges. This behaviour of the density matrix implies essentially nonthermal state of the system. Below we shall probe how coherent are bosonic modes and  irradiation in this state. At larger values of $\Omega$ the density matrix again becomes diagonal.  This is the end of the narrow line fluorescence regime.

More informative is the Wigner function evolution  presented in Figs.~\ref{Distr} (d)-(f) for the same driving amplitudes. At very small driving fields the Wigner function has only one peak located near the state with zero quanta of bosonic mode. Above the threshhold, Fig.~\ref{Distr} (h), it has the ``atoll'' form and the blue hole corresponds to negative values of the Wigner function. This is known for dissipative JCM under coherent excitation of bosonic mode.~\cite{Dombi2014JPB}  But here we consider the generalised Hamiltonian where also coherent driving of TLS is taken into account.  Then we see new behavior: the Wigner function first splits up into two narrow peaks (narrow line fluorescence regime), Fig.~\ref{Distr}(j), but going to larger field amplitudes we observe that the left peak of the Wigner function disappears and only the right one survives, Fig.~\ref{Distr} (k). This behaviour should correspond to the end of narrow line fluorescence regime.

Finally we consider the  spectra of the bosonic correlators, $S(t)=\langle{\hat a^ + }(t+\tau) {\hat a}(t)\rangle$.  Graphs in  Fig.~\ref{spectrum} explicitly illustrate the presence of the lower and upper threshold of the domain where the linewidth do reduces.

Generally speaking there are a number of known cases when noise may effectively increase coherence: this is the stochastic resonance~\cite{gammaitoni1998stochastic,wellens2004stochastic}. Metamolecule is strongly correlated system and it is not clear in advance what noise  may do. Possible interpretation of the  narrowing of fluorescence line is the quantum stochastic resonance.  However it can be a kind of lasing as well: We have the threshold and the population inversion in our system. We see that spectral line, which length is determined by the level of spontaneous emission, is narrowing. This points out that coherence the system output signal increases. Another point is splitting and narrowing of the Wigner function. The characteristic width of the Wigner function determines quantum fluctuations. In the regime of the developed generation the same situation takes place. In Ref.~\onlinecite{chow2014emission} three lasing criteria have been formulated: 1) transition to lasing should be accompanied by nonlinear dependence of  the boson number on external drive, 2)  emission linewidth should strongly decrease in the lasing regime and 3) the second order coherence function $g^{(2)}(0)$ should tend to one which indicates the coherent output.  In our model fluorescence in the regime of  ``narrow spectral line'' shows all these properties.

Also we note that the metamolecule we focus on has many features typical for recent nanolaser type experiments where the bosonic mode is the plasmonic mode of metallic nanoparticle while semiconductor quantum dot or dye molecule play the role of the active medium and similar narrowing of spectral line and threshold behavior have been seen~\cite{noginov2009demonstration,oulton2009Nature,lu2012Science}. In these experiments pulse pumping has been used.  However the usual duration of the pump pulse was about 10~ns. While all characteristic times of nanolaser/spaser like problems are much shorter (the nanosecond scale).  So this pulse pumping can be well approximated by coherent pumping like we do.
Recently another experiment showed the fluorescence under coherent pumping identified with lasing~\cite{khajavikhan2012Nature}.


To conclude, we describe the behavior of metamolcule driven by the external field. It is shown that fluorescence at dissipation larger than driving is accompanied by narrowing of the spectral line by the order of magnitude and  splitting of the Wigner function of the emitted light. This effect may find applications in quantum systems where  strong noise due to high losses is unavoidable problem like in plasmonics.

We thank A. Vinogradov and Yu. Lozovik for helpful discussions. N.C. acknowledges for the hospitality Laboratoire de Physique Théorique, Toulouse and CNRS where this work was finalized. The work was partially funded by Dynasty and Russian Foundation for Basic Research No.~13-02-00407, while supercomputer simulations were funded by Russian Scientific Foundation №14-12-01185. We thank Supercomputer Centers of Russian Academy of Sciences and National research Center Kurchatov Institute for access to URAL, JSCC and HPC supercomputer clusters.

\bibliographystyle{osajnl}
\bibliography{refs}

\end{document}